\documentstyle[12pt,aasms4]{article}

\begin{document}
\Large 
\centerline{\bf A smoother end to the dark ages}
\normalsize 
\author{\small Zolt\'{a}n Haiman}

\begin{verbatim}
                   News and Views, Nature, 7 April 2011
\end{verbatim}

\vskip 0.2in 

\hrule 

\vskip 0.1in 

\noindent
{\em\large Independent lines of evidence suggest that the first stars,
which ended the cosmic dark ages, came in pairs, rather than
singly. This could change the prevailing view that the early Universe
had a Swiss--cheese--like appearance.}

\vskip 0.2in 

\hrule 

\vskip 0.2in 

\thispagestyle{empty}

After a spectacular birth, our Universe quickly became a dull place,
with the glow of the Big Bang fading away and the first stars and
galaxies yet to appear. These cosmic dark ages lasted for a hundred
million years. According to a growing body of evidence, the latest of
which is described by Mirabel el al.\cite{Mirabel} in a paper
published in Astronomy \& Astrophysics, many of the first stars that
put an end to the dark ages may have formed in pairs.  The appearance
of the first stars marked a significant milestone, separating the
history of the Universe into two stages. The first stage is well
understood: dark matter, primordial ionized plasma and radiation
formed a nearly uniform mixture, expanding and cooling continuously
with cosmic time. When the temperature of the plasma dropped below
3,000 Kelvin, neutral hydrogen and helium atoms formed
everywhere. Spatial variations in the density and temperature of the
plasma were initially minuscule, but gravitational instability
amplified these over time, and allowed the collapse of dense gaseous
structures.

In the second stage, stars lit up inside these structures, and started
wreaking havoc. The radiation of the stars penetrated the neutral
cosmic plasma, once again ionizing and heating it, and modifying the
formation of the subsequent generations of stars. At the time of the
transition between the two stages, the 100--million--year-old Universe
may have resembled Swiss cheese: cold and neutral background gas was
filled with numerous roughly spherical, hot, ionized holes surrounding
the sites where the earliest stars had lit up. There is, however, a
different possibility, in which energetic X--ray radiation -- not
normally associated with stars -- is present during the
transition. The evidence that the first stars may have formed in pairs
makes the latter hypothesis more likely.  Over the past decade, a
theoretical paradigm has emerged in which the first stars formed in
isolation and were about 100 times more massive than a typical
present--day star, such as our own Sun. The first dense gaseous
structures in which stars formed were very small; they had a total
mass of the order of a million solar masses — approximately a million
times smaller than a typical present--day galaxy such as the Milky
Way. Molecular hydrogen, which formed efficiently in these dense
regions, allowed the gas to radiate efficiently and lose its pressure
support, making it collapse further\cite{HTL96}. Three--dimensional
simulations\cite{Bromm09,Abel02} revealed that these structures form
at the intersections of thin filaments, forming a cosmic web--like
structure. They have also shown that a small fraction of the gas --
about 100 solar masses -- flows along the filaments coherently toward
the central region of each such dense knot, without any sign of
fragmenting into pieces. However, recent
simulations\cite{Turk09,Stacy10,Prieto11,Greif11}, of higher
resolution than the earlier simulations\cite{Bromm09,Abel02}, have
suggested that the gas in the central regions does, eventually,
fragment into two or more distinct clumps, raising the possibility
that the first stars formed in pairs, or even in higher--multiple
systems.

Why would the companionship of the first stars matter for the rest of
the Universe?  As argued by Mirabel and colleagues\cite{Mirabel}, a
natural outcome of the latter hypothesis is for one member in a pair
of massive stars to implode and leave behind a black hole, remaining
gravitationally bound to its massive partner. The black hole can then
pull material off the surface of its unfortunate partner, and swallow
it efficiently. While devouring its partner, the black hole returns a
fraction of the ingested energy in the form of copious amounts of
X--rays. In fact, there are compelling examples of such
‘micro--quasars’ in the local Universe.  They appear more common in
smaller galaxies, as well as in galaxies whose chemical composition is
closer to the that of the pristine hydrogen + helium plasma in the
early universe, unpolluted by the heavier atoms produced by subsequent
generations of stars. The extrapolation of these local observations
suggests that such binaries were more common in the earliest, small
and primitive micro--galaxies.

If the majority of the first stars formed such binaries, they could
have produced sufficient X--rays to significantly change the
prevailing Swiss--cheese hypothesis. This possibility has been raised
in the past\cite{Oh01,Venkatesan01,Glover03,Chen04}, but is now worth
considering more seriously, in light of the new theoretical and
observational evidence. Unlike the ultraviolet ionizing radiation from
normal stars, X--rays with the right energy -- of order of a
kiloelectronvolt (keV) -- will travel, in the early Universe, across
vast distances, ionizing and heating the plasma much more
uniformly. If X--rays are sufficiently prevalent, a range of other
interesting effects will occur: the extra heating will raise the
pressure of the plasma everywhere, making it resistant to clumping,
and more difficult to compress to form new galaxies\cite{Oh01}.

On the other hand, X--rays can penetrate the successfully collapsing
galaxies and can ionize hydrogen and helium atoms in their
interior. This will catalyze the formation of molecular hydrogen, and
help the gas to cool and form new stars\cite{har00}. These effects
will leave behind their signatures in the spatial distribution of
neutral and ionized hydrogen and helium in the Universe. Mapping these
distributions — by measuring the 21--centimetre radio emission from
neutral hydrogen\cite{FOB06}, and the scattering of cosmic microwave
background radiation (relic radiation from the Big Bang) by free
electrons, or by examining the absorption spectra of distant galaxies
— is feasible in forthcoming experiments, and forms a major goal of
modern cosmology.

There are other possible sources of X--rays connected to the formation
of the first stars, for example gas accretion onto the black--hole
remnants left behind by the collapse of single
stars\cite{Madau04,Ricotti05}. Another possible source is supernovae
(SNe): if the first stars exploded as SNe, then similar X--rays would
be produced by thermal emission from the gas heated by these SNe, and
by the collisions between the energetic electrons produced in the SN
explosion and the cosmic microwave background
photons\cite{Oh01}. However, if micro--quasars were indeed as common,
and as efficient producers of X--ray radiation, as Mirabel and
colleagues argue1, they may well have dominated the X--ray production
in the transition epoch, when the first stars started to shine in the
Universe. They would then have been responsible for ending the dark
ages in a smooth fashion. The hardest X--ray photons (with energies
above a few keV) would be reaching us on Earth now, forming a feeble
X--ray background. Existing measurements place an upper limit on the
present--day value of this background, which is consistent with this
hypothesis\cite{DHL04}. The possibility of X--ray production by binary
stars should prompt further theoretical modelling of the population of
such binaries, including their abundance, radiation output and
spectra, as well as of the possible observable signatures they left
behind.

\small

\noindent{\it Zolt\'an Haiman is in the Department of Astronomy,\\ Columbia
 University, New York, NY 10027, USA.\\email: zoltan@astro.columbia.edu}

\end{document}